%% file: index_tracking_TRex.tex
\documentclass[journal]{IEEEtran}

\input{packages.tex}
\input{macros.tex}

\mathtoolsset{showonlyrefs=true} 
\restylefloat{table}

\begin{document}

\title{FDR-Controlled Portfolio Optimization for\\ Sparse Financial Index Tracking}

\author{
Jasin Machkour, 
Daniel P. Palomar, 
and Michael Muma 
\thanks{J. Machkour,~Student Member,~IEEE, and M. Muma, Senior Member,~IEEE, are with the Robust Data Science Group at Technische Universit\"at Darmstadt, Germany (e-mail: jasin.machkour@tu-darmstadt.de; michael.muma@tu-darmstadt.de). D. P. Palomar,~Fellow,~IEEE, is with the Convex Optimization Group, The Hong Kong University of Science and Technology, Hong Kong SAR, China (e-mail: palomar@ust.hk).}%
\thanks{The first author has been supported by the LOEWE initiative (Hesse, Germany) within the emergenCITY center. The second author has been supported by the Hong Kong GRF 16206123 research grant. The third author has been supported by the ERC Starting Grant ScReeningData (Project Number: 101042407).}%
\thanks{Extensive computations on the Lichtenberg high-performance computer of the Technische Universität Darmstadt were conducted for this research.}%
}%

\maketitle

\begin{abstract}
In high-dimensional data analysis, such as financial index tracking or biomedical applications, it is crucial to select the few relevant variables while maintaining control over the false discovery rate (FDR). In these applications, strong dependencies often exist among the variables (e.g., stock returns), which can undermine the FDR control property of existing methods like the model-X knockoff method or the T-Rex selector. To address this issue, we have expanded the T-Rex framework to accommodate overlapping groups of highly correlated variables. This is achieved by integrating a nearest neighbors penalization mechanism into the framework, which provably controls the FDR at the user-defined target level. A real-world example of sparse index tracking demonstrates the proposed method's ability to accurately track the S\&P 500 index over the past 20 years based on a small number of stocks. An open-source implementation is provided within the R package TRexSelector on CRAN.
\end{abstract}
\section{Introduction}
\label{sec: Introduction}
Financial index tracking is fundamental for the design of asset portfolios that are used to create exchange traded funds (ETFs) and hedging strategies of mutual funds~\cite{benidis2017sparse,prigent2007portfolio}. Prevalent index tracking approaches replicate an entire index (e.g., SPDR S\&P 500 ETF) by creating and regularly updating a full tracking portfolio. However, this leads to high transaction costs because it requires the regular purchase and disposition of all assets in an index. Therefore, sparse index tracking methods, which use a small fraction of the stocks that constitute an index, have been proposed~\cite{jansen2002optimal,maringer2007index,scozzari2013exact,xu2016efficient,benidis2017sparse}. The common disadvantage of existing sparse approaches is that they require the investor to choose the size of the tracking portfolio or the value of a sparsity tuning parameter. Since there exist no optimal strategies for the choice of these parameters, the authors of the aforementioned approaches resort to experimental choices or rules-of-thumb that often lead to sub-optimal tracking portfolios.

In this work, we propose an approach that automatically determines the size and composition of a sparse index tracking portfolio by controlling the false discovery rate (FDR)~\cite{benjamini1995controlling}. The FDR is the expected fraction of false discoveries (i.e., stocks that are irrelevant for tracking an index) among all selected stocks. The target FDR between $0$ and $100$\% expresses the level of the investor's willingness to sacrifice a small amount of transaction costs (arising from the inclusion of a few irrelevant stocks into the tracking portfolio) in order to obtain a diversified and yet small tracking portfolio.

Unfortunately, existing FDR-controlling methods are not suitable for sparse financial index tracking, since they are
\begin{enumerate}
\item not applicable in high-dimensional settings (e.g.,\cite{benjamini1995controlling,benjamini2001control,barber2015controlling}),
\item lose the FDR control property in the presence of groups of highly correlated variables (e.g.,~\cite{candes2018panning,machkour2021terminating,machkour2023ScreenTRex,scheidt2023FDRControlLaptop}), or
\item do not account for strongly overlapping groups of highly correlated variables which are characteristic for stock returns data (e.g.,~\cite{machkour2023InformedEN,machkour2024dependency}).
\end{enumerate}

Therefore, we propose a new FDR-controlling index tracking method by
\begin{enumerate}
\item extending the dependency-aware \textit{T-Rex} selector~\cite{machkour2024dependency} to account for strongly overlapping groups of highly correlated variables using a nearest neighbors penalization mechanism,
\item proving that it controls the FDR at the investor-specified target level, and
\item demonstrating its unique capability of accurately tracking the S\&P $500$ index over the last $20$ years based on a sparse, FDR-controlled, and quarterly updated portfolio.
\end{enumerate}
While this paper highlights an application in finance, the method more generally applies to any setting where the variables exhibit an overlapping groups dependency structure.

\textit{Organization}: Section~\ref{sec: The Dependency-Aware T-Rex Selector} briefly revisits the dependency-aware \textit{T-Rex} framework. Sections~\ref{sec: Proposed: FDR-Controlled Index Tracking} and~\ref{sec: Tracking the SP 500 Index} introduce the proposed FDR-controlling index tracking method and compare the tracking performance against benchmark methods.

\section{The Dependency-Aware T-Rex Selector}
\label{sec: The Dependency-Aware T-Rex Selector}
The ordinary \textit{T-Rex} selector~\cite{machkour2021terminating,machkour2023ScreenTRex} is a fast FDR-controlling variable selection method for high-dimensional data, where the number of variables $p$ may exceed the number of samples $n$. Its output is an FDR-controlled estimate of the true active variable set $\A \subseteq \lbrace 1, \ldots, p \rbrace$ whose cardinality is denoted by $| \A |$. The general methodology of the \textit{T-Rex} selector is depicted in Figure~\ref{fig: T-Rex selector framework}. First, $K \cdot L$ dummy predictor vectors of dimension $n$ are sampled from the univariate standard normal distribution and stacked columnwise into $K$ different dummy predictor matrices $\XK_{k} \in \mathbb{R}^{n \times L}$, $k = 1, \ldots, K$. These are then appended to the original predictor matrix $\X \in \mathbb{R}^{n \times p}$, i.e., $\XWK_{k} \coloneqq \big[ \X \,\, \XK_{k} \big]$. The enlarged predictor matrices $\XWK_{k}$ and the response vector $\y \in \mathbb{R}^{n}$ are the inputs of $K$ random experiments. In each of the independent experiments, a forward variable selection algorithm, such as the \textit{LARS}~\cite{efron2004least}, \textit{Lasso}~\cite{tibshirani1996regression}, or related methods (e.g.,~\cite{zou2005regularization,zou2006adaptive}), includes one variable in each forward selection iteration of the algorithm. Each random experiment is early terminated after $T \in \lbrace 1, \ldots, L \rbrace$ dummies are included along the forward selection path. The obtained candidate sets $\C_{k, L}(T)$, $k = 1, \ldots, K$, are used to compute the relative occurrences of the variables $j = 1, \ldots, p$, i.e.,
\begin{equation}
\Phi_{T,L}(j) \coloneqq 
\begin{cases}
\dfrac{1}{K} \sum\limits_{k = 1}^{K} \mathbbm{1}_{k}(j, T, L), & T \geq 1 \\
0, \quad & T = 0
\end{cases},
\label{eq: relative occurrences}
\end{equation}
where the indicator function $\mathbbm{1}_{k}(j, T, L)$ is equal to one if the $j$th variable is contained in the candidate set $\C_{k, L}(T)$ and zero otherwise. Finally, the set of selected variables $\widehat{\A}_{L}(v, T)$ consists of all variables whose relative occurrences exceed a voting threshold $v \in [ 0.5, 1 )$, i.e.,
\begin{equation}
\widehat{\A}_{L}(v, T) \coloneqq \lbrace j : \Phi_{T, L}(j) > v \rbrace.
\label{eq: selected active set}
\end{equation}
The extended calibration algorithm proposed in~\cite{machkour2021terminating} automatically determines a solution $(v^{*}, T^{*}, L^{*}) \in [ 0.5, 1 ) \times \lbrace 1, \ldots, L^{*} \rbrace \times \mathbb{N}_{+}$, such that the FDR is controlled at the user-specified target level $\alpha \in [ 0, 1 ]$, while maximizing the true positive rate (TPR). The FDR is the expected value of the false discovery proportion (FDP), i.e.,
\begin{equation}
\FDR \coloneqq \mathbb{E} \big[ \FDP \big] \coloneqq \mathbb{E} \bigg[ \dfrac{V_{T, L}(v)}{\max \big\lbrace 1, R_{T, L}(v) \big\rbrace} \bigg ]  \leq \alpha.
\label{eq: FDR controlled at target level}
\end{equation}
Here,
\begin{align}
& V_{T, L}(v) \coloneqq \big| \lbrace \text{null } j : \Phi_{T, L}(j) > v \rbrace \big|,
\label{eq: number of selected null variables}
\\
& R_{T, L}(v) = \big| \widehat{\A}_{L}(v, T) \big| \coloneqq \big| \lbrace j : \Phi_{T, L}(j) > v \rbrace \big|
\label{eq: number of selected variables}
\end{align}
denote, respectively, the number of selected null variables (i.e., false discoveries) and the total number of selected variables. The TPR is the expected value of the true positive proportion (TPP), i.e.,
\begin{equation}
\TPR \coloneqq \mathbb{E} \big[ \TPP \big] \coloneqq \mathbb{E} \bigg[ \dfrac{S_{T, L}(v)}{\max \big\lbrace 1, | \A | \big\rbrace} \bigg ],
\label{eq: FDR controlled at target level}
\end{equation}
where $S_{T, L}(v) \coloneqq \big| \lbrace \text{active } j : \Phi_{T, L}(j) > v \rbrace \big|$ denotes the number of selected true active variables. 
%
\input{T_Rex_framework.tex}

%

The dependency-aware \textit{T-Rex} (\textit{T-Rex+DA}) selector~\cite{machkour2024dependency} is an extension of the \textit{T-Rex} selector that accounts for dependencies among variables by replacing the ordinary relative occurrences in~\eqref{eq: relative occurrences} by dependency-aware relative occurrences
\begin{equation}
\Phi_{T, L}^{\DA}(j, \rho_{\thr}) \coloneqq \Psi_{T, L}(j, \rho_{\thr}) \cdot \Phi_{T, L}(j),
\label{eq: dependency-aware relative occurrences}
\end{equation}
where the penalty factor $\Psi_{T, L}(j, \rho_{\thr}) \in [0.5, 1]$ is defined by
\begin{align}
&\Psi_{T, L}(j, \rho_{\thr}) \coloneqq
\label{eq: dependency factor}
\\
&
\begin{cases}
 \dfrac{1}{2 \,\,\, - \,\,\, \smashoperator{\min\limits_{j^{\prime} \in \Gr(j, \rho_{\thr})}} \quad\, \big\lbrace \big| \Phi_{T, L}(j) - \Phi_{T, L}(j^{\prime}) \big| \big\rbrace}, &\!\! \Gr(j, \rho_{\thr}) \neq \varnothing
 \\[1em]
 1 / 2, &\!\! \Gr(j, \rho_{\thr}) = \varnothing
 \end{cases}.
\end{align}
$\Psi_{T, L}(j, \rho_{\thr})$ penalizes the $j$th ordinary relative occurrence according to its resemblance with other ordinary relative occurrences in its associated variable group $\Gr(j, \rho_{\thr}) \subseteq \lbrace 1, \ldots, p \rbrace$. The parameter $\rho_{\thr}$ controls the size of the variable groups and is automatically determined by the \textit{T-Rex+DA} calibration algorithm such that the FDR is controlled. In~\cite{machkour2024dependency}, it has been proposed to use a hierarchical clustering approach to design disjoint variable groups that allow for FDR control in the presence of disjoint groups of highly correlated variables.

\section{Proposed: FDR-Controlled Index Tracking}
\label{sec: Proposed: FDR-Controlled Index Tracking}
The existing design of $\Gr(j, \rho_{\thr})$ for the dependency-aware \textit{T-Rex} framework~\cite{machkour2024dependency} does not allow for arbitrary overlapping groups of highly correlated variables, which are characteristic for stock returns data. Therefore, in this section, we propose a new design for the variable groups $\Gr(j, \rho_{\thr})$, $j = 1, \ldots p$, which incorporates a nearest neighbors (NN) penalization mechanism into the dependency-aware \textit{T-Rex} framework. We then prove that the proposed approach controls the FDR at the user-specified target level. Finally, the stock returns data model is introduced and the proposed FDR-controlled index tracking algorithm is formulated.

\subsection{Group Design and FDR Control}
\label{subsec: Group Design and FDR Control}
For the proposed NN group design, each variable $j$ is assigned a variable group that contains variables whose correlations with variable $j$ exceed a threshold  $\rho_{\thr} \in [0, 1]$, i.e.,
\begin{align}
\Gr(j, \rho_{\thr}) \coloneqq \big\lbrace j^{\prime} \in \lbrace 1, &\ldots, p \rbrace \backslash \big\lbrace j \big\rbrace : 
\\
&| \corr(\x_{j}, \x_{j^{\prime}})| \geq \rho_{\thr} \big\rbrace.
\label{eq: nearest neighbors group design principle}
\end{align}
Clearly, in contrast to~\cite{machkour2024dependency}, this group design allows for overlapping groups of highly correlated variables. The challenge now is to jointly determine $\rho_{\thr}$ in~\eqref{eq: nearest neighbors group design principle} and the other parameters of the dependency-aware \textit{T-Rex} selector (i.e., $v$, $T$, and $L$) such that the proposed NN group design yields FDR-controlled solutions 
\begin{equation}
\widehat{\A}_{L}(v, \rho_{\thr}, T) \coloneqq \lbrace j : \Phi_{T, L}^{\NN}(j, \rho_{\thr}) > v \rbrace,
\label{eq: T-Rex+DA+NN selected active set}
\end{equation}
where
\begin{equation}
\Phi_{T, L}^{\NN}(j, \rho_{\thr}) \coloneqq \Psi_{T, L}^{\NN}(j, \rho_{\thr}) \cdot \Phi_{T, L}(j)
\label{eq: dependency-aware relative occurrences with nearest neighbors group design}
\end{equation}
is the dependency-aware relative occurrence of the $j$th variable using the proposed NN group design in~\eqref{eq: nearest neighbors group design principle}.

Note that the distinction of the penalty functions $\Psi_{T, L}$ and $\Psi_{T, L}^{\NN}$ in~\eqref{eq: dependency factor} and~\eqref{eq: dependency-aware relative occurrences with nearest neighbors group design} stems from their underlying group designs. The underlying group design of $\Psi_{T, L}$ only allows for disjoint variable groups. In contrast, the proposed NN group design in~\eqref{eq: nearest neighbors group design principle} extends the scope of the \textit{T-Rex+DA} selector by allowing for overlapping groups of highly correlated variables, which are characteristic for applications such as the considered financial index tracking.

In order to prove that the proposed NN group design controls the FDR, we first define the conservative FDP estimator\footnote{Conservative is meant in the sense that $\mathbb{E}[ \FDP ] \leq \mathbb{E}[ \widehat{\FDP} ]$~\cite{machkour2021terminating}.}
\begin{equation}
\widehat{\FDP}(v, \rho_{\thr}, T, L) \coloneqq \dfrac{\widehat{V}_{T, L}(v, \rho_{\thr})}{\max \big\lbrace 1, R_{T, L}(v, \rho_{\thr}) \big\rbrace},
\label{eq: FDP estimator}
\end{equation}
where $R_{T, L}(v, \rho_{\thr}) \coloneqq | \widehat{\A}_{L}(v, \rho_{\thr}, T) |$ is the number of selected variables. Second, let $\widehat{\A}(v, \rho_{\thr}) \coloneqq \widehat{\A}_{L}(v, \rho_{\thr}, T)$ and $\Delta\Phi_{t, L}^{\NN}(j, \rho_{\thr}) \coloneqq \Phi_{t, L}^{\NN}(j, \rho_{\thr}) - \Phi_{t - 1, L}^{\NN}(j, \rho_{\thr})$. Then, the proposed estimator of the unknown number of selected null variables $V_{T, L}(v, \rho_{\thr})$ in~\eqref{eq: FDP estimator} is defined by
\begingroup
\allowdisplaybreaks
\begin{align}
&\widehat{V}_{T, L}(v, \rho_{\thr}) \coloneqq
\sum\limits_{j \in \widehat{\A}(v, \rho_{\thr})} \Big( 1 - \Phi_{T, L}^{\NN}(j, \rho_{\thr}) \Big)
\label{eq: estimator of V_T_L with nearest neighbors group design}
\\
&+ \underbrace{\sum\limits_{t = 1}^{T} \dfrac{p - \sum_{q = 1}^{p}\Phi_{t, L}^{\NN}(q, \rho_{\thr})}{L - (t - 1)} \cdot \dfrac{\sum\limits_{j \in \widehat{\A}(v, \rho_{\thr})} \Delta\Phi_{t, L}^{\NN}(j, \rho_{\thr})}{\sum\limits_{j \in \widehat{\A}(0.5, \rho_{\thr})} \Delta\Phi_{t, L}^{\NN}(j, \rho_{\thr})}}_{\eqqcolon \widehat{V}_{T, L}^{\prime}(v, \rho_{\thr})},
\end{align}
\endgroup
which is similar to the estimator that was defined in~\cite{machkour2021terminating,machkour2024dependency} with the innovation that the ordinary relative occurrences are replaced by the proposed relative occurrences in~\eqref{eq: dependency-aware relative occurrences with nearest neighbors group design}.

With all necessary definitions in place, we now state and prove that any quadruple $(v, \rho_{\thr}, T, L) \in [0.5, 1) \times [0, 1] \times \lbrace 1, \ldots, L \rbrace \times \mathbb{N}_{+}$, for which the conservative FDP estimator in~\eqref{eq: FDP estimator} does not exceed the user-specified target level $\alpha \in [0, 1]$, yields FDR control. To simultaneously maximize the number of selected variables (to obtain highest possible TPR while controlling FDR), we choose the lowest possible voting threshold $v$ that maintains FDR control, i.e.,
\begin{equation}
v \coloneqq \inf \lbrace \nu \in [0.5, 1) : \widehat{\FDP}(\nu, \rho_{\thr}, T, L) \leq \alpha \rbrace.
\label{eq: v stopping time}
\end{equation}
\begin{thm}[FDR control]
Let $\Gr(j, \rho_{\thr})$ be as defined in~\eqref{eq: nearest neighbors group design principle} and $K \rightarrow \infty$. Suppose that $\widehat{V}_{T, L}^{\prime}(v, \rho_{\thr}) > 0$. Then, for any quadruple $(v, \rho_{\thr}, T, L) \in [0.5, 1) \times [0, 1] \times \lbrace 1, \ldots, L \rbrace \times \mathbb{N}_{+}$ that satisfies Eq.~\eqref{eq: v stopping time}, it holds that $\FDR(v, \rho_{\thr}, T, L) \leq \alpha$, i.e., FDR control at the user-specified target level $\alpha \in [0, 1]$.
\label{theorem: FDR control nearest neighbors group design}
\end{thm}

\begingroup
\allowdisplaybreaks
\begin{proof}
An upper bound on the FDR is given by
\begin{align}
\FDP(v, \rho_{\thr}, T, L) &= \dfrac{V_{T, L}(v, \rho_{\thr})}{\max \big\lbrace 1, R_{T, L}(v, \rho_{\thr}) \big\rbrace}
\label{eq: proof - theorem - FDR control nearest neighbors group design 1}
\\
&= \widehat{\FDP}(v, \rho_{\thr}, T, L) \cdot \dfrac{V_{T, L}(v, \rho_{\thr})}{\widehat{V}_{T, L}(v, \rho_{\thr})}
\label{eq: proof - theorem - FDR control nearest neighbors group design 2}
\\
&\leq \alpha \cdot \dfrac{V_{T, L}(v, \rho_{\thr})}{\widehat{V}_{T, L}(v, \rho_{\thr})},
\label{eq: proof - theorem - FDR control nearest neighbors group design 3}
\\
&\leq \alpha \cdot \dfrac{V_{T, L}(v, \rho_{\thr})}{\widehat{V}_{T, L}^{\prime}(v, \rho_{\thr})},
\label{eq: proof - theorem - FDR control nearest neighbors group design 3}
\end{align}
where the second, third, and fourth line follow from~\eqref{eq: FDP estimator}, ~\eqref{eq: v stopping time}, and~\eqref{eq: estimator of V_T_L with nearest neighbors group design}, respectively. Taking the expected value yields
\begin{equation}
\FDR(v, \rho_{\thr}, T, L) \leq \alpha \cdot \mathbb{E} \bigg[ \dfrac{V_{T, L}(v, \rho_{\thr})}{\widehat{V}_{T, L}^{\prime}(v, \rho_{\thr})} \bigg].
\label{eq: proof - theorem - FDR control nearest neighbors group design 5}
\end{equation}
It remains to prove that $\mathbb{E}[ V_{T, L}(v, \rho_{\thr}) / \widehat{V}_{T, L}^{\prime}(v, \rho_{\thr}) ] \leq 1$. Along the lines of the proof of Lemma~1 in~\cite{machkour2024dependency}, it can be shown that $V_{T, L}(v, \rho_{\thr}) / \widehat{V}_{T, L}^{\prime}(v, \rho_{\thr})$ is a backward-running super-martingale with respect to the filtration $\mathcal{F}_{v} \coloneqq \sigma(\lbrace V_{T, L}(v, \rho_{\thr}) \rbrace_{u \geq v}, \lbrace \widehat{V}_{T, L}^{\prime}(v, \rho_{\thr}) \rbrace_{u \geq v})$. Therefore, and since $v$ in~\eqref{eq: v stopping time} is $\mathcal{F}_{v}$-measurable, an upper bound on $\mathbb{E}[ V_{T, L}(v, \rho_{\thr}) / \widehat{V}_{T, L}^{\prime}(v, \rho_{\thr}) ]$ can be derived using Doob's optional stopping theorem~\cite{williams1991probability}. First, using the definition of $V_{T, L}(v)$ in~\eqref{eq: number of selected null variables} and defining $\widehat{V}_{T, L}^{\prime}(0.5) \coloneqq \sum_{t = 1}^{T} \frac{p - \sum_{q = 1}^{p} \Phi_{t, L}(q)}{L - (t - 1)}$, we obtain
\begin{align}
\mathbb{E} \bigg[ \dfrac{V_{T, L}(v, \rho_{\thr})}{\widehat{V}_{T, L}^{\prime}(v, \rho_{\thr})} \bigg] &\leq \mathbb{E} \bigg[ \dfrac{V_{T, L}(0.5, \rho_{\thr})}{\widehat{V}_{T, L}^{\prime}(0.5, \rho_{\thr})} \bigg]
\label{eq: proof - theorem - FDR control nearest neighbors group design 6}
\\
&\leq \mathbb{E} \bigg[ \dfrac{V_{T, L}(0.5)}{\widehat{V}_{T, L}^{\prime}(0.5)} \bigg]
\label{eq: proof - theorem - FDR control nearest neighbors group design 7}
\\
&\leq 1.
\label{eq: proof - theorem - FDR control nearest neighbors group design 8}
\end{align}
Inequality~\eqref{eq: proof - theorem - FDR control nearest neighbors group design 6} follows from Doob's optional stopping theorem. Inequality~\eqref{eq: proof - theorem - FDR control nearest neighbors group design 7} follows from the following two inequalities:
\begin{align}
\text{(i)  } V_{T, L}(0.5, \rho_{\thr}) &= \lbrace \text{null } j : \Phi_{T, L}^{\NN}(j, \rho_{\thr}) > 0.5 \rbrace
\label{eq: proof - theorem - FDR control nearest neighbors group design 9}
\\
&\leq \lbrace \text{null } j : \Phi_{T, L}(j) > 0.5 \rbrace = V_{T, L}(0.5),
\label{eq: proof - theorem - FDR control nearest neighbors group design 11}
\\[-1em]
\text{(ii)  } \widehat{V}_{T, L}^{\prime}(0.5, \rho_{\thr}) &= \sum_{t = 1}^{T} \dfrac{p - \sum_{q = 1}^{p} \Phi_{t, L}^{\NN}(q, \rho_{\thr})}{L - (t - 1)}
\label{eq: proof - theorem - FDR control nearest neighbors group design 12}
\\
&\geq \sum_{t = 1}^{T} \dfrac{p - \sum_{q = 1}^{p} \Phi_{t, L}(q)}{L - (t - 1)} = \widehat{V}_{T, L}^{\prime}(0.5).
\label{eq: proof - theorem - FDR control nearest neighbors group design 14}
\end{align}
where the inequalities in (i) and (ii) both follow from Eq.~\eqref{eq: dependency-aware relative occurrences with nearest neighbors group design} and the fact that $\Psi_{T, L}^{\NN}(j, \rho_{\thr}) \leq 1$. Finally, inequality~\eqref{eq: proof - theorem - FDR control nearest neighbors group design 8} has already been proven to hold (see proof of Theorem~1 in~\cite{machkour2021terminating}) and, thus, it holds that $\FDR(v, \rho_{\thr}, T, L) \leq \alpha$.
\label{proof: theorem - FDR control nearest neighbors group design}
\end{proof}
\endgroup
\input{sp500_index_tracking.tex}

\subsection{Stock Returns Data Model}
\label{subsec: Stock Returns Data Model}
As suggested in~\cite{benidis2017sparse}, we model the stocks returns as a linear regression model $\y = \X\w + \bepsilon$, where $\w = [ w_{1} \cdots w_{p} ]^{\top} \in \mathbb{R}^{p}$ is the asset weight vector and $\bepsilon = [ \epsilon_{1} \cdots \epsilon_{n} ]^{\top} \in \mathbb{R}^{n}$ is an additive noise vector. Here, $\y = [ y_{1} \cdots y_{n} ]^{\top} \in \mathbb{R}^{n}$ is the daily index returns vector, i.e., $y_{i} = (\ind_{i} - \ind_{i - 1}) / \ind_{i - 1}$, $i = 1, \ldots, n$, where $\ind_{i}$ is the closing price of the index on day $i$ and $\ind_{0}$ is the closing price at the first day of the considered period. Analogously, $\X = [\x_{1} \cdots \x_{p}] \in \mathbb{R}^{n \times p}$ is the matrix containing the daily returns of the stocks $\x_{j} = [ x_{1, j} \cdots x_{n, j} ]^{\top} \in \mathbb{R}^{n}$, i.e., $x_{i, j} = (\price_{i, j} - \price_{i - 1, j}) / \price_{i - 1, j}$, $i = 1, \ldots, n$, where $\price_{i, j}$ is the closing price of the $j$th stock on day $i$ and $\price_{0, j}$ is the closing price on the first day of the period.

\subsection{Algorithm: FDR-Controlled Index Tracking}
\label{subsec: Algorithm: FDR-Controlled Index Tracking}
The general goal of sparse index tracking is to determine a sparse estimator of the asset weight vector $\w$ that tracks the index $\y$ sufficiently well using few relevant assets while obeying the following two rules~\cite{benidis2017sparse}:
\begin{enumerate}[label=\arabic*., ref=\arabic*]
\itemsep0em
\item Shorting stocks is not allowed, i.e., $w_{j} \geq 0$, $j = 1, \ldots, p$.
\item The available budget has to be invested, i.e., $\| \w \|_{1} = 1$.
\end{enumerate}
The proposed algorithm regularly updates the FDR-controlled tracking portfolio in a rolling-window fashion, as suggested in~\cite{benidis2017sparse}, while satisfying the ``no-shorting'' and budget constraint in every training period $m \in \lbrace 1, \ldots, M \rbrace$. Therefore, in every training period $m$, we first run the dependency-aware \textit{T-Rex} selector with the proposed nearest neighbors group design in~\eqref{eq: nearest neighbors group design principle} and then solve the constrained quadratic problem
\begin{align}
\min_{\w_{m}} \, & \| \y_{m} - \X_{m, \widehat{\A}^{(m)}} \cdot \w_{m} \|_{2}^{2} + \lambda \| \w_{m} \|_{2}^{2} \\
&\text{subject to} \quad \w_{m} \geq \boldsymbol{0} \,\, \text{and} \,\, \| \w_{m} \|_{1} = 1.
\label{eq: QP for no-shorting and budget constraint}
\end{align}
Here, $\X_{m, \widehat{\A}^{(m)}}$ contains only the daily returns of the selected subset of stocks in $ \widehat{\A}^{(m)}$. Algorithm~\ref{algorithm: FDR-controlled index tracking} summarizes the proposed index tracking method. An open-source implementation of the proposed and related methods~\cite{koka2024false,machkour2024sparse} is available within the R package `TRexSelector' on CRAN~\cite{machkour2022TRexSelector}.

\section{Tracking the S\&P 500 Index}
\label{sec: Tracking the SP 500 Index}
We consider the S\&P $500$ index in the $20$ year period from $01/01/2003 - 29/09/2023$, which gives us $5{,}220$ trading days (i.e., samples) in total. These are divided into $M = 86$ training and $86$ testing periods, where the first period can only be used for training and the last period only for testing. Using the same rolling window approach as in~\cite{benidis2017sparse}, the portfolio is updated quarterly, i.e., all training and testing periods consist of $n = 60$ trading days. After removing all stocks that contain missing values, $p = 390$ candidate stocks are left.
\begin{algorithm}[t]
\caption{FDR-controlled index tracking.}
\begin{alglist}
\item \textbf{Input}: $\alpha \in [0, 1]$, $\X$, $\y$.
\label{algorithm: FDR-controlled index tracking Step 1}
\item \textbf{For} $m = 1, \ldots, M$ \textbf{do}:
\begin{enumerate}[label*=\arabic*.]
\item[2.1.] \textbf{Compute} $\Gr(j, \rho_{\thr})$, $j = 1, \ldots, p$, as defined in~\eqref{eq: nearest neighbors group design principle} for all  $\rho_{\thr} \in \lbrace 0, 0.01, 0.02, \ldots, 1 \rbrace$.
\item[2.2.] \textbf{Run} the dependency-aware \textit{T-Rex} selector (Algorithm~1 in~\cite{machkour2024dependency}) with the proposed groups $\Gr(j, \rho_{\thr})$ in~\eqref{eq: nearest neighbors group design principle} to obtain an optimal solution $(v^{*}, \rho_{\thr}^{*}, T^{*}, L^{*})$ and the corresponding FDR-controlled set of selected stocks $\widehat{\A}^{(m)} \coloneqq \widehat{\A}_{L^{*}}^{(m)}(v^{*}, \rho_{\thr}^{*}, T^{*}, L^{*})$.
\item[2.3.] \textbf{Solve} the quadratic optimization problem in~\eqref{eq: QP for no-shorting and budget constraint}.
\end{enumerate}
\label{algorithm: FDR-controlled index tracking Step 2}
\item \textbf{Output}: Portfolio weight vectors for each quarter, i.e., $\widehat{\w}_{1}, \ldots, \widehat{\w}_{M}$.
\label{algorithm: FDR-controlled index tracking Step 3}
\end{alglist}
\label{algorithm: FDR-controlled index tracking}
\end{algorithm}

As FDR-controlling benchmark methods, we consider the ordinary \textit{T-Rex} selector~\cite{machkour2021terminating} and the \textit{model-X} knockoff+ method~\cite{candes2018panning} for the variable selection step in Algorithm~\ref{algorithm: FDR-controlled index tracking}. We also consider the non-FDR-controlling state-of-the-art sparse index tracking method \textit{ALAIT - ETE}~\cite{benidis2017sparse} that solves a Lasso-type optimization problem with the same no-shorting and budget constraints using a majorization minimization approach. The target FDR $\alpha$ is set to $30$\% and the sparsity tuning parameter for \textit{ALAIT - ETE} to $\lambda = 10^{-7}$, as suggested in~\cite{benidis2017sparse,benidis2019sparseIndexTracking}. The tracking performance is measured by the more comprehensible wealth. That is, we do not consider the absolute value of the index but set the value of the index at the start of the tracking period to one and the wealth represents how an investors wealth has changed with respect to the reference. A mathematical definition of the wealth is given in~\cite{benidis2017sparse}. Note that index tracking does not aim at achieving the highest possible wealth but at an accurate tracking of the wealth corresponding to the index.

The results in Figure~\ref{fig: sp500_index_tracking} and Table~\ref{table: results S&P 500 tracking} show that the proposed dependency-aware \textit{T-Rex} selector with the nearest neighbors penalization mechanism (\textit{T-Rex+DA+NN}) has the best index tracking performance, while requiring the smallest number of stocks in almost all quarters.

\section{Conclusion}
\label{sec: Conclusion}
The dependency-aware \textit{T-Rex} framework has been extended to account for overlapping groups of highly correlated variables, which are characteristic for stock returns data. The proposed method provably controls the FDR and has been successfully applied in tracking the S\&P $500$ index over the last $20$ years with higher accuracy and fewer stocks compared to state-of-the-art methods. The proposed method does not require sparsity parameter tuning. Therefore, it is a promising approach for index tracking as well as other applications where overlapping groups of highly correlated variables exist.

\vfill\cleardoublepage

\bibliographystyle{IEEEtran_newLineInBibURL}
\IEEEtriggeratref{14}
\bibliography{bibliography}

\vfill\cleardoublepage

\typeout{get arXiv to do 4 passes: Label(s) may have changed. Rerun}

\end{document}

%% file: packages.tex
\usepackage[T1]{fontenc}
\usepackage[utf8]{inputenc}
\usepackage[ngerman,english]{babel}
\usepackage{amsmath, amssymb, ﻿amsfonts}   
\usepackage{cuted} 
\usepackage{graphicx}         
\usepackage{algorithmic}
\usepackage{algorithm}
\usepackage{stfloats}
\usepackage{url}
\usepackage{verbatim}
\usepackage{graphicx}
\usepackage{cite}

\usepackage{upgreek}			  
\usepackage{mathtools}		  
\usepackage{url}					
\usepackage{enumitem}		
\usepackage{accents}			
\usepackage{bbm}
\usepackage{xcolor}
\usepackage{lipsum}
\usepackage{dirtytalk}
\usepackage{xcolor}
\usepackage{float}
\usepackage{accents}
\usepackage[export]{adjustbox}
\usepackage{setspace}

	\usepackage[caption=false,font=normalsize,labelfont=sf,textfont=sf]{subfig}

\usepackage{pifont} 
\usepackage{caption}

\usepackage{booktabs, tabularx}
\usepackage{array}

\usepackage{amsthm}
\usepackage{enumitem}

\usepackage{tikz}
\usetikzlibrary{decorations.markings,chains,matrix}
\usetikzlibrary{intersections,calc}
\usepackage{tikz-3dplot} 
\usepackage{pgfplots}
\usepackage{pgfplotstable}
\pgfplotsset{compat = newest}
\usepackage{datatool}
\usepackage{textcomp}
\usetikzlibrary{shapes,arrows}
\usetikzlibrary{decorations.pathreplacing,angles,quotes}

%% file: macros.tex
\newtheorem{thm}{Theorem}

\DeclareMathOperator{\FDP}{FDP}
\DeclareMathOperator{\FDR}{FDR}
\DeclareMathOperator{\TPP}{TPP}
\DeclareMathOperator{\TPR}{TPR}

\DeclareMathOperator{\thr}{thr}

\DeclareMathOperator{\DA}{DA}

\DeclareMathOperator{\ind}{index}
\DeclareMathOperator{\price}{price}
\DeclareMathOperator{\Gr}{Gr}

\DeclareMathOperator{\corr}{corr}

\DeclareMathOperator{\NN}{NN}

\newtheorem{innercustomgeneric}{\customgenericname}
\providecommand{\customgenericname}{}
\newcommand{\newcustomtheorem}[2]{%
  \newenvironment{#1}[1]
  {%
   \renewcommand\customgenericname{#2}%
   \renewcommand\theinnercustomgeneric{##1}%
   \innercustomgeneric
  }
  {\endinnercustomgeneric}
}

\newcustomtheorem{customthm}{Theorem}
\newcustomtheorem{customlemma}{Lemma}
\newcustomtheorem{customcor}{Corollary}
\newcustomtheorem{customassumption}{Assumption}

\newcommand{\y}{\boldsymbol{y}}
\newcommand{\x}{\boldsymbol{x}}

\newcommand{\X}{\boldsymbol{X}}

\newcommand{\bepsilon}{\boldsymbol{\epsilon}}

\newcommand{\A}{\mathcal{A}}

\newcommand{\XK}{\boldsymbol{\protect\accentset{\circ}{X}}}

\newcommand{\XWK}{\boldsymbol{\widetilde{X}}}

\newcommand{\C}{\mathcal{C}}

\newcommand{\w}{\boldsymbol{w}}

\makeatletter
\let\save@mathaccent\mathaccent
\newcommand*\if@single[3]{%
  \setbox0\hbox{${\mathaccent"0362{#1}}^H$}%
  \setbox2\hbox{${\mathaccent"0362{\kern0pt#1}}^H$}%
  \ifdim\ht0=\ht2 #3\else #2\fi
  }
\newcommand*\rel@kern[1]{\kern#1\dimexpr\macc@kerna}
\newcommand*\widebar[1]{\@ifnextchar^{{\wide@bar{#1}{0}}}{\wide@bar{#1}{1}}}
\newcommand*\wide@bar[2]{\if@single{#1}{\wide@bar@{#1}{#2}{1}}{\wide@bar@{#1}{#2}{2}}}
\newcommand*\wide@bar@[3]{%
  \begingroup
  \def\mathaccent##1##2{%
    \let\mathaccent\save@mathaccent
    \if#32 \let\macc@nucleus\first@char \fi
    \setbox\z@\hbox{$\macc@style{\macc@nucleus}_{}$}%
    \setbox\tw@\hbox{$\macc@style{\macc@nucleus}{}_{}$}%
    \dimen@\wd\tw@
    \advance\dimen@-\wd\z@
    \divide\dimen@ 3
    \@tempdima\wd\tw@
    \advance\@tempdima-\scriptspace
    \divide\@tempdima 10
    \advance\dimen@-\@tempdima
    \ifdim\dimen@>\z@ \dimen@0pt\fi
    \rel@kern{0.6}\kern-\dimen@
    \if#31
      \overline{\rel@kern{-0.6}\kern\dimen@\macc@nucleus\rel@kern{0.4}\kern\dimen@}%
      \advance\dimen@0.4\dimexpr\macc@kerna
      \let\final@kern#2%
      \ifdim\dimen@<\z@ \let\final@kern1\fi
      \if\final@kern1 \kern-\dimen@\fi
    \else
      \overline{\rel@kern{-0.6}\kern\dimen@#1}%
    \fi
  }%
  \macc@depth\@ne
  \let\math@bgroup\@empty \let\math@egroup\macc@set@skewchar
  \mathsurround\z@ \frozen@everymath{\mathgroup\macc@group\relax}%
  \macc@set@skewchar\relax
  \let\mathaccentV\macc@nested@a
  \if#31
    \macc@nested@a\relax111{#1}%
  \else
    \def\gobble@till@marker##1\endmarker{}%
    \futurelet\first@char\gobble@till@marker#1\endmarker
    \ifcat\noexpand\first@char A\else
      \def\first@char{}%
    \fi
    \macc@nested@a\relax111{\first@char}%
  \fi
  \endgroup
}
\makeatother

\DeclareFontFamily{U}{dutchcal}{\skewchar\font=45}
\DeclareFontShape{U}{dutchcal}{m}{n}{<-> s*[1.2] dutchcal-r}{}
\DeclareFontShape{U}{dutchcal}{b}{n}{<-> s*[1.2] dutchcal-b}{}
\DeclareMathAlphabet{\mathdutchcal}{U}{dutchcal}{m}{n}
\SetMathAlphabet{\mathdutchcal}{bold}{U}{dutchcal}{b}{n}
\DeclareMathAlphabet{\mathdutchbcal}{U}{dutchcal}{b}{n}

\newlist{steps}{enumerate}{1}
\setlist[steps, 1]{label = {Step \arabic*:}, ref = {Step \arabic*}}

\newlist{alglist}{enumerate}{1}
\setlist[alglist, 1]{label = {\arabic*.}, ref = {\arabic*}}

\newcolumntype{L}[1]{>{\raggedright\arraybackslash}p{#1}}
\newcolumntype{C}[1]{>{\centering\arraybackslash}p{#1}}
\newcolumntype{R}[1]{>{\raggedleft\arraybackslash}p{#1}}

\definecolor{dark_green}{RGB}{102,166,30}
\definecolor{applegreen}{rgb}{0.55, 0.71, 0.0}

\definecolor{dark_red}{RGB}{217,95,2}
\definecolor{bittersweet}{rgb}{1.0, 0.44, 0.37}

\definecolor{dark_yellow}{RGB}{230,171,2}
\definecolor{bananayellow}{rgb}{1.0, 0.88, 0.21}

%% file: T_Rex_framework.tex
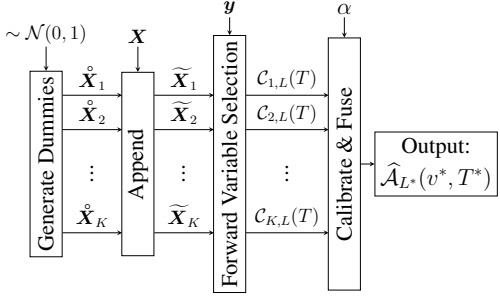
\begin{figure}[t]
\begin{center}
\scalebox{0.61}{
\begin{tikzpicture}[>=stealth]

  \coordinate (orig)   at (0,0);
  \coordinate (sample)   at (1,0.5);
  \coordinate (merge)   at (3,0.5);
  \coordinate (varSelect)   at (5,0.5);
  \coordinate (tFDR)   at (15.5,-1.1);
  \coordinate (fuse)   at (7.5,0.5);
  \coordinate (output)   at (9.5,0.5);
  
  \coordinate (between_scale_rank)   at (0.5,0.31);
  \coordinate (X_prime_to_tFDR_point)   at (0.5,6);
  \coordinate (X_prime_to_tFDR_point_point)   at (9,6);
  \coordinate (X_prime_to_merge_point)   at (0.5,-2.5);
  \coordinate (center_to_tFDR_point)   at (5.00,3.7);
  \coordinate (tFDR_to_fuse_point)   at (7.5,3.7);
  \coordinate (tFDR_to_sample_point)   at (4,5);

  \coordinate (Arrow_N_GenDummy)   at (1,3.06);
  \coordinate (Arrow_X_indVar)   at (3,3.06);
  \coordinate (Arrow_targetFDR_tFDR)   at (10,5.9);
  
   \coordinate (inference_Arrow)   at (15,-1.06);
   \coordinate (fuse_Arrow)   at (16.1,2.06);
  
  \coordinate (vdots1)   at (2.0,0.4);
  \coordinate (vdots2)   at (4.0,0.4);
  \coordinate (vdots3)   at (6.25,0.4);
  
  \coordinate (fuse_node)   at (15.00,0.5);
  
  \node[draw, minimum width=.7cm, minimum height=4cm, anchor=center , align=center] (C) at (sample) {\rotatebox{90}{\Large Generate Dummies}};
  \node[draw, minimum width=.7cm, minimum height=4cm, anchor=center, align=center] (D) at (merge) {\rotatebox{90}{\Large Append}};   
  \node[draw, minimum width=.7cm, minimum height=5.5cm, anchor=center, align=center] (E) at (varSelect) {\rotatebox{90}{\Large Forward Variable Selection}};
  \node[draw, minimum width=.7cm, minimum height=5.5cm, anchor=center, align=center] (H) at (fuse) {\rotatebox{90}{\Large Calibrate \& Fuse}};
  \node[draw, minimum width=2.5cm, minimum height=.7cm, anchor=center, align=center] (N) at (output) {\Large Output: \\[0.3em] \Large $\widehat{\mathcal{A}}_{L^{*}}(v^{*}, T^{*})$};
  \node (J) at (vdots1) {\Large $\vdots$};
  \node (K) at (vdots2) {\Large $\vdots$};
  \node (L) at (vdots3) {\Large $\vdots$};
  
  \draw[->] (Arrow_N_GenDummy) -- node[above, pos = 0.1]{\large $\sim\mathcal{N}(0, 1)$} ($(C.90)$); 
  \draw[->] (Arrow_X_indVar) -- node[above, pos = 0.1]{\large $\X$} ($(D.90)$); 
     
  \draw[->] ($(C.0) + (0,1.5)$) -- node[above]{\large $\XK_{1}$} ($(D.0) + (-0.7,1.5)$);
  \draw[->] ($(C.0) + (0,0.75)$) -- node[above]{\large $\XK_{2}$} ($(D.0) + (-0.7,0.75)$);
  \draw[->] ($(C.0) + (0,-1.5)$) -- node[above]{\large $\XK_{K}$} ($(D.0) + (-0.7,-1.5)$);
     
  \draw[->] ($(D.0) + (0,1.5)$) -- node[above]{\large $\XWK_{1}$} ($(E.0) + (-0.7,1.5)$);
  \draw[->] ($(D.0) + (0,0.75)$) -- node[above]{\large $\XWK_{2}$} ($(E.0) + (-0.7,0.75)$);
  \draw[->] ($(D.0) + (0,-1.5)$) -- node[above]{\large $\XWK_{K}$} ($(E.0) + (-0.7,-1.5)$);
     
  \draw[->] ($(E.0) + (0,1.5)$) -- node[above]{\large $\C_{1, L}(T)$} ($(H.0) + (-0.7,1.5)$);
  \draw[->] ($(E.0) + (0,0.75)$) -- node[above]{\large $\C_{2, L}(T)$} ($(H.0) + (-0.7,0.75)$);
  \draw[->] ($(E.0) + (0,-1.5)$) -- node[above]{\large $\C_{K, L}(T)$} ($(H.0) +  (-0.7,-1.5)$);
     
  \draw[->] (center_to_tFDR_point) -- node[above, pos = 0.1]{\large $\y$} ($(E.90)$); 
  \draw[->] (tFDR_to_fuse_point) -- node[above, pos = 0.1]{\Large $\alpha$} ($(H.90)$);
 
  \coordinate (between_varSelect_fuse1)   at ($(E.0) + (2.75,1.5)$);
  \coordinate (between_varSelect_fuse2)   at ($(E.0) + (2.75,0.75)$);
  \coordinate (between_varSelect_fuse3)   at ($(E.0) + (2.75,-1.5)$);
 
  \draw[->] (H) -- (N);
\end{tikzpicture}}
\end{center}
\setlength{\abovecaptionskip}{2pt}
\setlength{\belowcaptionskip}{-3pt}
\caption{Simplified \textit{T-Rex} selector framework~\cite{machkour2021terminating,machkour2022TRexGVS}.}
\label{fig: T-Rex selector framework}
\end{figure}

%% file: sp500_index_tracking.tex
\begin{figure*}
\begin{minipage}[t][][b]{0.64\textwidth}
    \centering
   \scalebox{1.02}{
  			\includegraphics[width=0.99\linewidth]{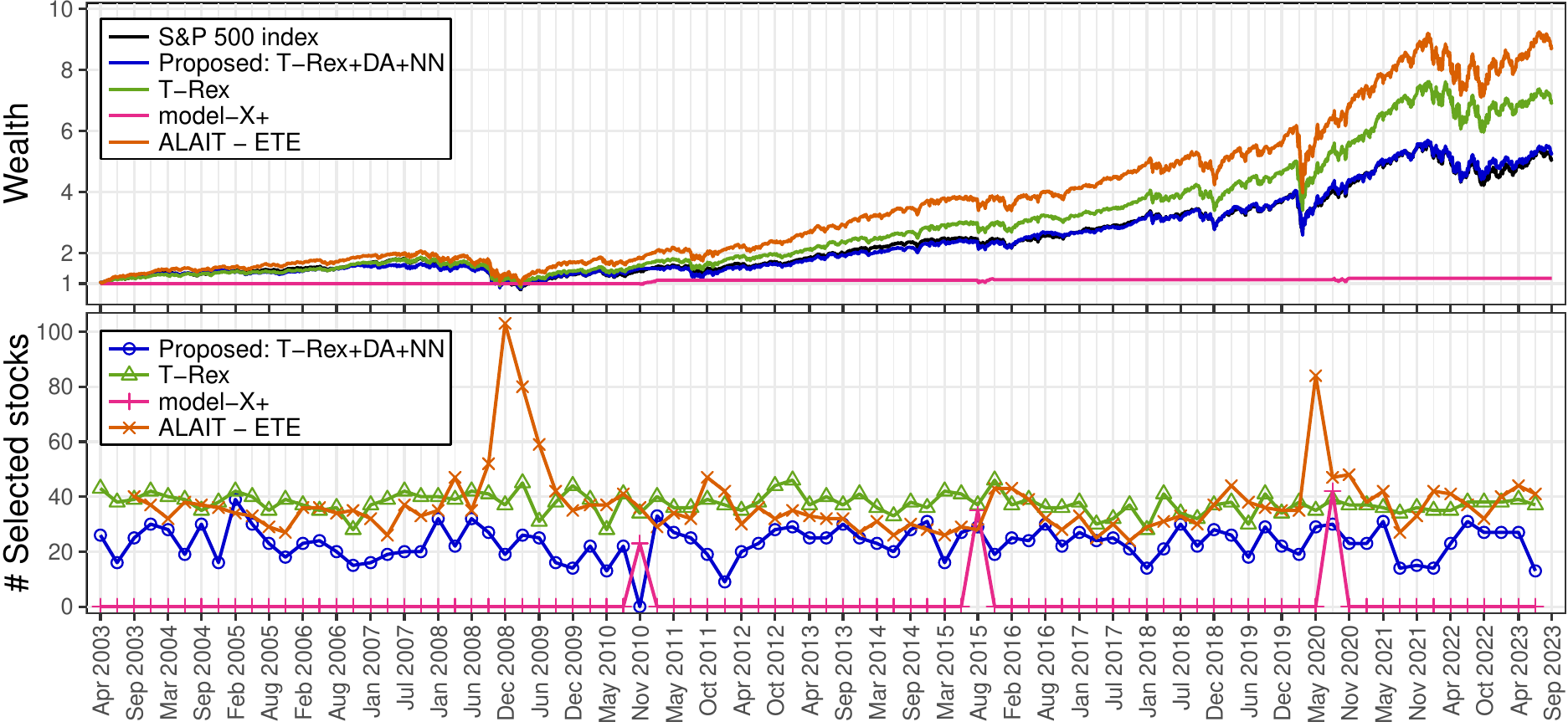}
  		}
  		\setlength{\abovecaptionskip}{-5pt}
  		\setlength{\belowcaptionskip}{-10pt}
    \captionof{figure}{The proposed \textit{T-Rex+DA+NN} selector closely follows the S\&P $500$ index using the fewest number of stocks in almost all quarters.}
     \label{fig: sp500_index_tracking}
  \end{minipage}
  \hfill
  \begin{minipage}[t][][b]{0.34\textwidth}
    \centering
    \scalebox{0.9}{
    \setlength{\tabcolsep}{2.8pt}
\begin{tabular}{
lccccr
}
\toprule
& &
\multicolumn{2}{c}{\bfseries \textit{ALAIT - ETE}} &
\multicolumn{2}{c}{\textit{\bfseries \textit{T-Rex+DA+NN}}}\\
\cmidrule(lr){3-4} \cmidrule(lr){5-6}
   $\lambda$ & $\alpha$ & \# stocks & MSTE &  \# stocks & MSTE\\
   \midrule
   $16.67$ & $20$ & $12.72$ & $0.96$ & $\boldsymbol{9.62}$ & $ \boldsymbol{0.91}$\\
   $6.67$ & $25$ & $18.73$ & $0.76$ & $\boldsymbol{13.67}$ & $\boldsymbol{0.03}$\\
   $3.33$ & $30$ & $24.23$ & $1.32$ & $\boldsymbol{19.72}$ & $\boldsymbol{0.22}$\\
\bottomrule
\end{tabular}
}
\setlength{\belowcaptionskip}{-10pt}
\captionof{table}{The proposed \textit{T-Rex+DA+NN} selector achieves a lower average (over all quarters) mean squared wealth tracking error (MSTE) using a smaller portfolio. Another major advantage is that the proposed method only requires choosing an interpretable target FDR level $\alpha$ (in \%), while the benchmark method relies on the uninterpretable tuning parameter $\lambda$ ($\times 10^{-7}$).}
      \label{table: results S&P 500 tracking}
    \end{minipage}
\end{figure*}

%% file: index_tracking_TRex.bbl
\begin{thebibliography}{10}
\providecommand{\url}[1]{#1}
\csname url@samestyle\endcsname
\providecommand{\newblock}{\relax}
\providecommand{\bibinfo}[2]{#2}
\providecommand{\BIBentrySTDinterwordspacing}{\spaceskip=0pt\relax}
\providecommand{\BIBentryALTinterwordstretchfactor}{4}
\providecommand{\BIBentryALTinterwordspacing}{\spaceskip=\fontdimen2\font plus
\BIBentryALTinterwordstretchfactor\fontdimen3\font minus
  \fontdimen4\font\relax}
\providecommand{\BIBforeignlanguage}[2]{{%
\expandafter\ifx\csname l@#1\endcsname\relax
\typeout{** WARNING: IEEEtran.bst: No hyphenation pattern has been}%
\typeout{** loaded for the language `#1'. Using the pattern for}%
\typeout{** the default language instead.}%
\else
\language=\csname l@#1\endcsname
\fi
#2}}
\providecommand{\BIBdecl}{\relax}
\BIBdecl

\bibitem{benidis2017sparse}
K.~Benidis, Y.~Feng, and D.~P. Palomar, ``Sparse portfolios for
  high-dimensional financial index tracking,'' \emph{IEEE Trans. Signal
  Process.}, vol.~66, no.~1, pp. 155--170, 2017.

\bibitem{prigent2007portfolio}
J.-L. Prigent, \emph{Portfolio optimization and performance analysis}.\hskip
  1em plus 0.5em minus 0.4em\relax CRC Press, 2007.

\bibitem{jansen2002optimal}
R.~Jansen and R.~Van~Dijk, ``Optimal benchmark tracking with small
  portfolios,'' \emph{J. Portf. Manag.}, vol.~28, no.~2, p.~33, 2002.

\bibitem{maringer2007index}
D.~Maringer and O.~Oyewumi, ``Index tracking with constrained portfolios,''
  \emph{Intell. Syst. Account. Finance Manag.}, vol.~15, no. 1-2, pp. 57--71,
  2007.

\bibitem{scozzari2013exact}
A.~Scozzari, F.~Tardella, S.~Paterlini, and T.~Krink, ``Exact and heuristic
  approaches for the index tracking problem with {UCITS} constraints,''
  \emph{Ann. Oper. Res.}, vol. 205, pp. 235--250, 2013.

\bibitem{xu2016efficient}
F.~Xu, Z.~Lu, and Z.~Xu, ``An efficient optimization approach for a
  cardinality-constrained index tracking problem,'' \emph{Optim. Methods
  Softw.}, vol.~31, no.~2, pp. 258--271, 2016.

\bibitem{benjamini1995controlling}
Y.~Benjamini and Y.~Hochberg, ``Controlling the false discovery rate: a
  practical and powerful approach to multiple testing,'' \emph{J. R. Stat. Soc.
  Ser. B. Stat. Methodol.}, vol.~57, no.~1, pp. 289--300, 1995.

\bibitem{benjamini2001control}
Y.~Benjamini and D.~Yekutieli, ``The control of the false discovery rate in
  multiple testing under dependency,'' \emph{Ann. Statist.}, vol.~29, no.~4,
  pp. 1165--1188, 2001.

\bibitem{barber2015controlling}
R.~F. Barber and E.~J. Cand{\`e}s, ``Controlling the false discovery rate via
  knockoffs,'' \emph{Ann. Statist.}, vol.~43, no.~5, pp. 2055--2085, 2015.

\bibitem{candes2018panning}
E.~J. Cand{\`e}s, Y.~Fan, L.~Janson, and J.~Lv, ``Panning for gold:
  ‘model-{X}’ knockoffs for high dimensional controlled variable
  selection,'' \emph{J. R. Stat. Soc. Ser. B. Stat. Methodol.}, vol.~80, no.~3,
  pp. 551--577, 2018.

\bibitem{machkour2021terminating}
J.~Machkour, M.~Muma, and D.~P. Palomar, ``The terminating-random experiments
  selector: Fast high-dimensional variable selection with false discovery rate
  control,'' \emph{arXiv preprint arXiv:2110.06048}, 2022.

\bibitem{machkour2023ScreenTRex}
------, ``False discovery rate control for fast screening of large-scale
  genomics biobanks,'' in \emph{Proc. 22nd IEEE Statist. Signal Process.
  Workshop (SSP)}, 2023, pp. 666--670.

\bibitem{scheidt2023FDRControlLaptop}
F.~Scheidt, J.~Machkour, and M.~Muma, ``Solving {FDR}-controlled sparse
  regression problems with five million variables on a laptop,'' in \emph{Proc.
  IEEE 9th Int. Workshop Comput. Adv. Multi-Sensor Adapt. Process. (CAMSAP)},
  2023, pp. 116--120.

\bibitem{machkour2023InformedEN}
J.~Machkour, M.~Muma, and D.~P. Palomar, ``The informed elastic net for fast
  grouped variable selection and {FDR} control in genomics research,'' in
  \emph{Proc. IEEE 9th Int. Workshop Comput. Adv. Multi-Sensor Adapt. Process.
  (CAMSAP)}, 2023, pp. 466--470.

\bibitem{machkour2024dependency}
------, ``High-dimensional false discovery rate control for dependent
  variables,'' \emph{arXiv preprint arXiv:2401.15796}, 2024.

\bibitem{efron2004least}
B.~Efron, T.~Hastie, I.~Johnstone, and R.~Tibshirani, ``Least angle
  regression,'' \emph{Ann. Statist.}, vol.~32, no.~2, pp. 407--499, 2004.

\bibitem{tibshirani1996regression}
R.~Tibshirani, ``Regression shrinkage and selection via the lasso,'' \emph{J.
  R. Stat. Soc. Ser. B. Stat. Methodol.}, vol.~58, no.~1, pp. 267--288, 1996.

\bibitem{zou2005regularization}
H.~Zou and T.~Hastie, ``Regularization and variable selection via the elastic
  net,'' \emph{J. R. Stat. Soc. Ser. B. Stat. Methodol.}, vol.~67, no.~2, pp.
  301--320, 2005.

\bibitem{zou2006adaptive}
H.~Zou, ``The adaptive lasso and its oracle properties,'' \emph{J. Amer.
  Statist. Assoc.}, vol. 101, no. 476, pp. 1418--1429, 2006.

\bibitem{machkour2022TRexGVS}
J.~Machkour, M.~Muma, and D.~P. Palomar, ``False discovery rate control for
  grouped variable selection in high-dimensional linear models using the
  {T-Knock} filter,'' in \emph{30th Eur. Signal Process. Conf. (EUSIPCO)},
  2022, pp. 892--896.

\bibitem{williams1991probability}
D.~Williams, \emph{Probability with martingales}.\hskip 1em plus 0.5em minus
  0.4em\relax Cambridge Univ. Press, 1991.

\bibitem{koka2024false}
T.~Koka, J.~Machkour, and M.~Muma, ``False discovery rate control for
  {G}aussian graphical models via neighborhood screening,'' \emph{arXiv
  preprint arXiv:2401.09979}, 2024.

\bibitem{machkour2024sparse}
J.~Machkour, A.~Breloy, M.~Muma, D.~P. Palomar, and F.~Pascal, ``Sparse {PCA}
  with false discovery rate controlled variable selection,'' \emph{arXiv
  preprint arXiv:2401.08375}, 2024.

\bibitem{machkour2022TRexSelector}
\BIBentryALTinterwordspacing
J.~Machkour, S.~Tien, D.~P. Palomar, and M.~Muma, \emph{TRexSelector: T-Rex
  Selector: {H}igh-Dimensional Variable Selection \& FDR Control}, 2022, {R}
  package version 0.0.1. [Online].\par Available:
  \url{https://CRAN.R-project.org/package=TRexSelector}
\BIBentrySTDinterwordspacing

\bibitem{benidis2019sparseIndexTracking}
\BIBentryALTinterwordspacing
K.~Benidis and D.~P. Palomar, \emph{sparseIndexTracking: Design of portfolio of
  stocks to track an index}, 2019, {R} package version 0.1.1. [Online].\par
  Available: \url{https://CRAN.R-project.org/package=sparseIndexTracking}
\BIBentrySTDinterwordspacing

\end{thebibliography}
